\documentstyle[12pt,axodraw,epsfig]{article}

\def\beq{\begin{equation}}
\def\ra{\rightarrow}
\def\eeq{\end{equation}}

\def\bea{\begin{eqnarray}}
\def\eea{\end{eqnarray}}

\def\lam{\lambda}

\renewcommand{\bottomfraction}{1.0} 
\renewcommand{\textfraction}{0.0}   

\begin{document}

\renewcommand{\bottomfraction}{1.0} 
\renewcommand{\textfraction}{0.0}   

\title{\begin{flushright}
\vspace*{-1.2 cm}
{\normalsize hep-ph/0002220} \\
\vspace*{-0.4 cm}
{\normalsize CERN-TH/2000-060}  \\
\end{flushright} 
\vspace*{0.7 cm} {\Large \bf 
Three-body Supersymmetric Top Decays} \\
\author{{\bf Alexander Belyaev, John Ellis} and {\bf Smaragda Lola }\\
{\small CERN Theory Division, CH-1211 Geneva 23, Switzerland } \\ 
}}
\date{}
\maketitle
\vspace*{-0.5 cm}
{\bf Abstract:
{\small 
We discuss three-body supersymmetric top decays,
in schemes both with and without $R$-parity conservation,
assuming that sfermion masses are larger than $m_t$.
We find that MSSM top decays into chargino/neutralino
pairs have a strong kinematic suppression in the region of
the supersymmetric parameter space consistent with the LEP limits,
with a decay width $\leq 10^{-5}$~GeV. MSSM top decays into
neutralino pairs have less kinematical suppression, but require a
flavour-changing vertex, and are likely to have a smaller rate.
On the other hand, $R$-violating decays
to single charginos, neutralinos and conventional fermions can be larger
for values of the $R$-violating couplings still permitted by other
upper limits. The cascade decays of the charginos and neutralinos
may lead to spectacular signals with explicit lepton-number violation, such
as like-sign lepton events.
}}

\section{Introduction}

Future colliders such as LHC or an $e^+ e^-$ linear collider, e.g., TESLA,
will produce ${\bar t} t$ pairs with  high
statistics, and may be regarded as top factories.
For example, the cross section for $t\bar{t}$
production at the LHC is calculated to be of the order of
$800~pb$~\cite{Bonciani:1998vc}, corresponding to
the production of $1.6\times 10^7$ or $1.6\times 10^8$
top quarks per year in the  low- and high-luminosity regimes: 10 $fb^{-1}$
and 100 $fb^{-1}$, respectively.
This means, in principle, that if some rare top-decay channel
has a clean experimental signature without Standard Model
background, one could hope to measure the branching ratio 
with a sensitivity of $10^{-6}$ or $10^{-7}$.
The expected top production rate at TESLA  is lower
by a couple of orders of magnitude, but in this case the
events are much cleaner, with reduced backgrounds~\cite{topTESLA}.

The study of rare decays of the
top quark, the heaviest particle discovered so far,
may provide an interesting probe for physics beyond the Standard Model.
Indeed, novel top decays are predicted in many proposed extensions of the
Standard Model~\cite{Simmonds}, particularly in models of flavour
physics~\cite{FCC}.
In the minimal supersymmetric extension
of the Standard Model (MSSM) with conserved $R$ parity and
light sfermions and gauginos, two-body top decays 
into final states containing sparticle pairs
would be expected~\cite{2body1}. In supersymmetric models
with $R$ parity violated,
two-body decays of the top quark into a single sfermion may arise
if the sfermions are light~\cite{2body2}. Another possibility, even
in the case of heavier sfermions,is single-neutralino production
in top decays~\cite{3bodyneu}.

The purpose of this paper is to extend previous work to
a more complete study of three-body top decays.
We first study chargino-neutralino production
in the MSSM \cite{Sol}: 
$t \rightarrow \chi^+ \chi b$, and then
neutralino-pair production: $t \rightarrow \chi \chi c$, which could
in principle also arise, particularly if there is large 
$\tilde{c}-\tilde{t}$ mixing. Unfortunately, we find that
the present LEP constraints on the MSSM parameter space already
impose strong kinematic restrictions on $\Gamma(t
\rightarrow \chi^+ \chi b)$ decay, so that it is $\leq 10^{-5}$~GeV.
We find that $\Gamma(t \rightarrow \chi \chi c)$ is likely to be
even smaller for any amount of $\tilde{c}-\tilde{t}$ mixing.
A more interesting possibility in
models with $R$-violating supersymmetry is
single chargino and neutralino production:
$t \rightarrow \chi/\chi^+ {\bar q} q$. In view of
the very weak bounds on $R$-violating $t$-quark
couplings, we find that either of these rare top decays could
be observable at LHC and TESLA, failing which the
bounds on the corresponding couplings would be tightened.
Finally, novel top decays to three conventional fermions 
are also possible in models with $R$-violating $t$-quark
couplings, and might have distinctive signatures violating
lepton number.

\section{Top Decays to Charginos and Neutralinos in the MSSM}

\begin{figure}[h]
{\unitlength=1.5 pt
\SetScale{1.5}
\SetWidth{0.7}      
\begin{picture}(95,79)(0,0)
\Text(15.0,60.0)[r]{$t$}
\ArrowLine(16.0,60.0)(58.0,60.0) 
\Text(80.0,70.0)[l]{$b$}
\ArrowLine(58.0,60.0)(79.0,70.0) 
\Text(54.0,50.0)[r]{$W^+$}
\DashArrowLine(58.0,60.0)(58.0,40.0){3.0} 
\Text(80.0,50.0)[l]{$\chi$}
\Line(58.0,40.0)(79.0,50.0) 
\Text(80.0,30.0)[l]{$\chi^+$}
\ArrowLine(58.0,40.0)(79.0,30.0) 
\end{picture} \hspace*{-1cm}
\begin{picture}(95,79)(0,0)
\Text(15.0,60.0)[r]{$t$}
\ArrowLine(16.0,60.0)(58.0,60.0) 
\Text(80.0,70.0)[l]{$b$}
\ArrowLine(58.0,60.0)(79.0,70.0) 
\Text(54.0,50.0)[r]{$H^+$}
\DashArrowLine(58.0,60.0)(58.0,40.0){1.0} 
\Text(80.0,50.0)[l]{$\chi$}
\Line(58.0,40.0)(79.0,50.0) 
\Text(80.0,30.0)[l]{$\chi^+$}
\ArrowLine(58.0,40.0)(79.0,30.0) 
\end{picture}  \hspace*{-1cm}
\begin{picture}(95,79)(0,0)
\Text(15.0,60.0)[r]{$t$}
\ArrowLine(16.0,60.0)(58.0,60.0) 
\Text(80.0,70.0)[l]{$\chi$}
\Line(58.0,60.0)(79.0,70.0) 
\Text(54.0,50.0)[r]{$\tilde{t}_1$}
\DashArrowLine(58.0,60.0)(58.0,40.0){1.0} 
\Text(80.0,50.0)[l]{$\chi^+$}
\ArrowLine(58.0,40.0)(79.0,50.0) 
\Text(80.0,30.0)[l]{$b$}
\ArrowLine(58.0,40.0)(79.0,30.0) 
\end{picture} \\
\vspace*{-0.5cm}
\begin{picture}(95,79)(0,0)
\Text(15.0,60.0)[r]{$t$}
\ArrowLine(16.0,60.0)(58.0,60.0) 
\Text(80.0,70.0)[l]{$\chi$}
\Line(58.0,60.0)(79.0,70.0) 
\Text(54.0,50.0)[r]{$\tilde{t}_2$}
\DashArrowLine(58.0,60.0)(58.0,40.0){1.0} 
\Text(80.0,50.0)[l]{$\chi^+$}
\ArrowLine(58.0,40.0)(79.0,50.0) 
\Text(80.0,30.0)[l]{$b$}
\ArrowLine(58.0,40.0)(79.0,30.0) 
\end{picture}  \hspace*{-1cm}
\begin{picture}(95,79)(0,0)
\Text(15.0,60.0)[r]{$t$}
\ArrowLine(16.0,60.0)(58.0,60.0) 
\Text(80.0,70.0)[l]{$\chi^+$}
\ArrowLine(58.0,60.0)(79.0,70.0) 
\Text(54.0,50.0)[r]{$\tilde{b}_1$}
\DashArrowLine(58.0,60.0)(58.0,40.0){1.0} 
\Text(80.0,50.0)[l]{$\chi$}
\Line(58.0,40.0)(79.0,50.0) 
\Text(80.0,30.0)[l]{$b$}
\ArrowLine(58.0,40.0)(79.0,30.0) 
\end{picture}   \hspace*{-1cm}
\begin{picture}(95,79)(0,0)
\Text(15.0,60.0)[r]{$t$}
\ArrowLine(16.0,60.0)(58.0,60.0) \vspace*{-0.5cm}

\Text(80.0,70.0)[l]{$\chi^+$}
\ArrowLine(58.0,60.0)(79.0,70.0) 
\Text(54.0,50.0)[r]{$\tilde{b}_2$}
\DashArrowLine(58.0,60.0)(58.0,40.0){1.0} 
\Text(80.0,50.0)[l]{$\chi$}
\Line(58.0,40.0)(79.0,50.0) 
\Text(80.0,30.0)[l]{$b$}
\ArrowLine(58.0,40.0)(79.0,30.0) 
\end{picture} 
}
\vspace*{-0.5cm}
\caption{\label{fig1}
\em Three-body top decays to charginos and neutralinos
in the MSSM: $t \rightarrow \chi^+ \chi b$.}
\end{figure}
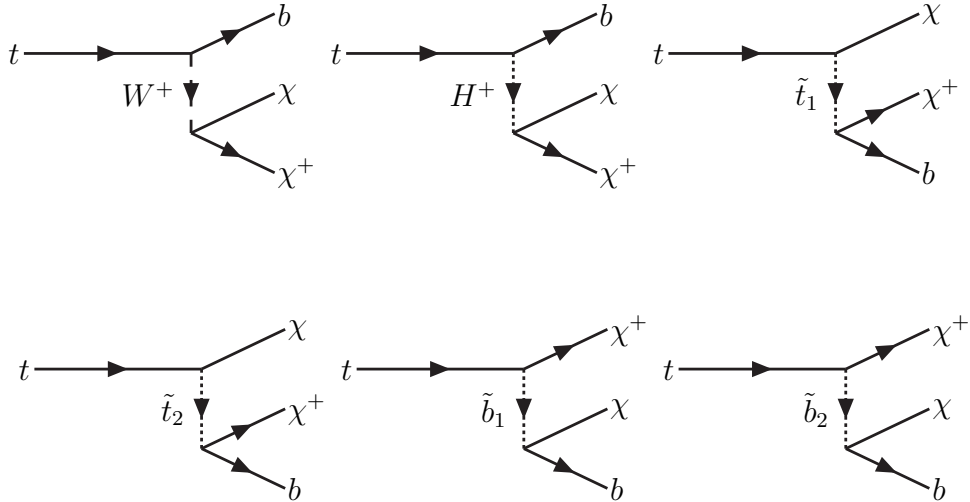

Two-body MSSM decays of the top quark would 
require the stop and/or sbottom quarks to be very light, which is
disfavoured by Tevatron data, so we focus here on three-body
MSSM top decays to light charginos $\chi^{\pm}$ and neutralinos
$\chi$. Even in the absence of $\tilde{c}-\tilde{t}$ mixing,
tops can decay to $\chi^+ \chi$ pairs via the diagrams shown in 
Fig.~\ref{fig1}.
The decay width for this process depends on:

$\bullet$ The magnitudes of the
neutralino and chargino masses and couplings,
which are functions of the MSSM
parameters. Assuming gaugino-mass unification,
the $U(1)$ and $SU(2)$ gaugino masses $M_1$ and
$M_2$  are related by $M_1 = \frac{5}{3}\tan\theta_W^2 M_2$,
in which case the gaugino and higgsino masses and couplings
are determined by $M_2$, $\mu$ and $\tan\beta$.

$\bullet$ The masses of the squarks $\tilde t$ and $\tilde b$.
We assume these to be kinematically inaccessible to $t$ decay, and
further require that the intermediate squarks be at least 10~GeV
off-shell.

\vspace*{0.2 cm}

\noindent
We recall that the chargino and
neutralino masses cannot be very light, in view
of the constraints imposed by LEP and other data.
Assuming gaugino-mass unification
and requiring the masses to be compatible with these constraints
significantly restricts the MSSM parameter space that can be of relevance.
These constraints also restrict the possible couplings of squarks
to charginos and neutralinos. For instance, in previous work
we found that there is a significant region of the MSSM parameter
space where the couplings of the $\tilde{q}_L$ to neutralinos are 
strongly suppressed due to accidental cancellations~\cite{Contour}, 
and the coupling of $\tilde{q}_R$ to neutralinos can be at most $\approx
0.23$. On the other hand, the couplings of squarks
to charginos are significantly larger, and can be as large as $\approx
0.5$ in regions of the parameter space where charginos and
neutralinos are relatively light.

We also note that there are regions of the MSSM parameter
space where the associated production of a chargino $\chi^+$ with the
second-lightest neutralino $\chi'$
may also be kinematically possible. In this case, however, the phase-space
suppression in models with gaugino-mass universality
is large enough to kill any amplification that might
arise due to the larger coupling of squarks to $\chi'$.
Finally, we point out that, if we postulate a deviation from gaugino-mass
universality, the models most likely to yield 
large rates would be those where $M_1$ is smaller
than $M_2$ at the input scale. In this case,
the correlation between chargino and neutralino masses is broken,
since the lightest neutralino is mainly a bino and
the lightest chargino is mainly a wino, and thus we can
allow for neutralino masses lower than the limit of about
45~GeV that is found in models with gaugino-mass universality.

\begin{figure}[h]
\vskip -0.5cm
\hspace*{-0.5cm}\epsfig{file=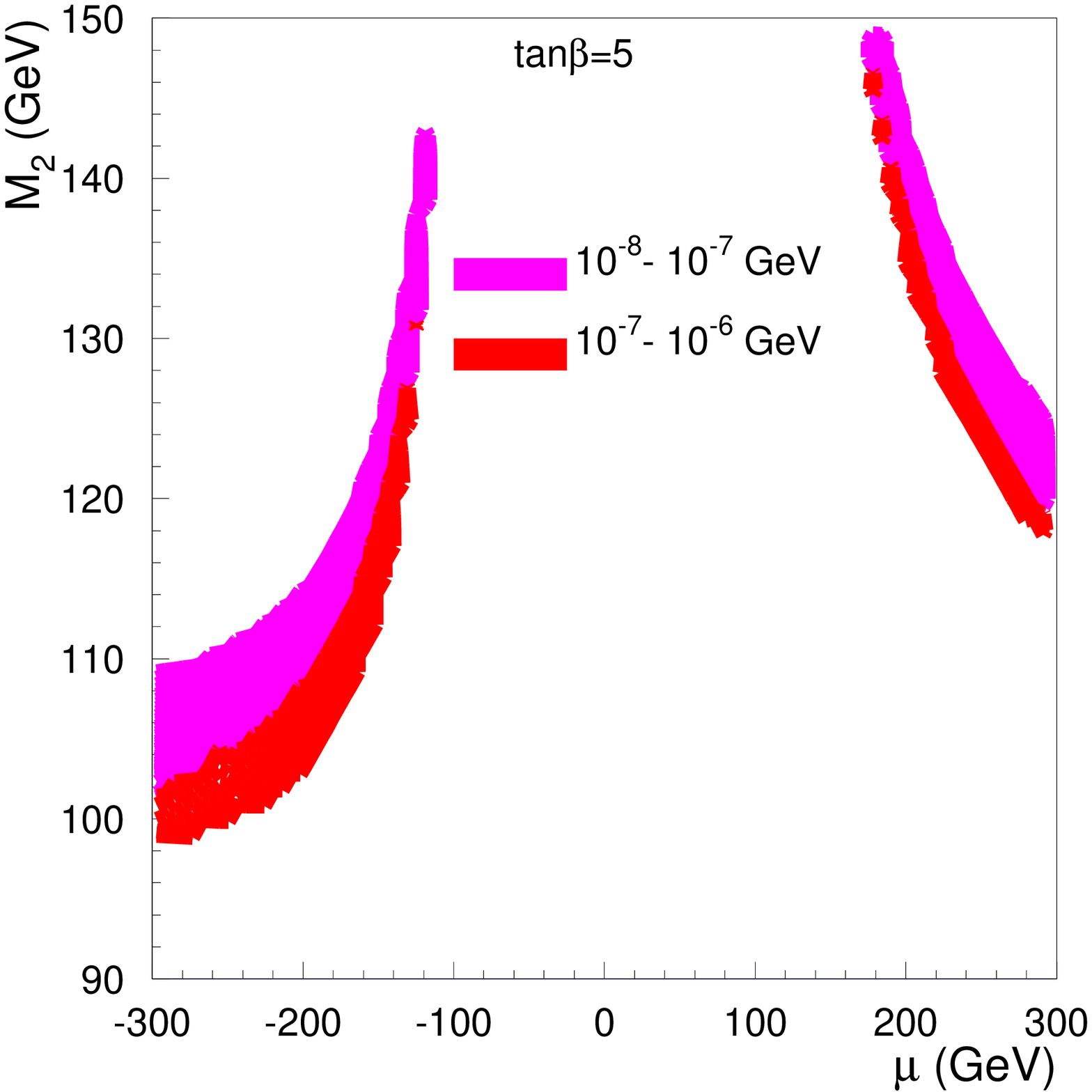,width=0.5\textwidth,height=0.5\textwidth}
\hspace*{-0.5cm}\epsfig{file=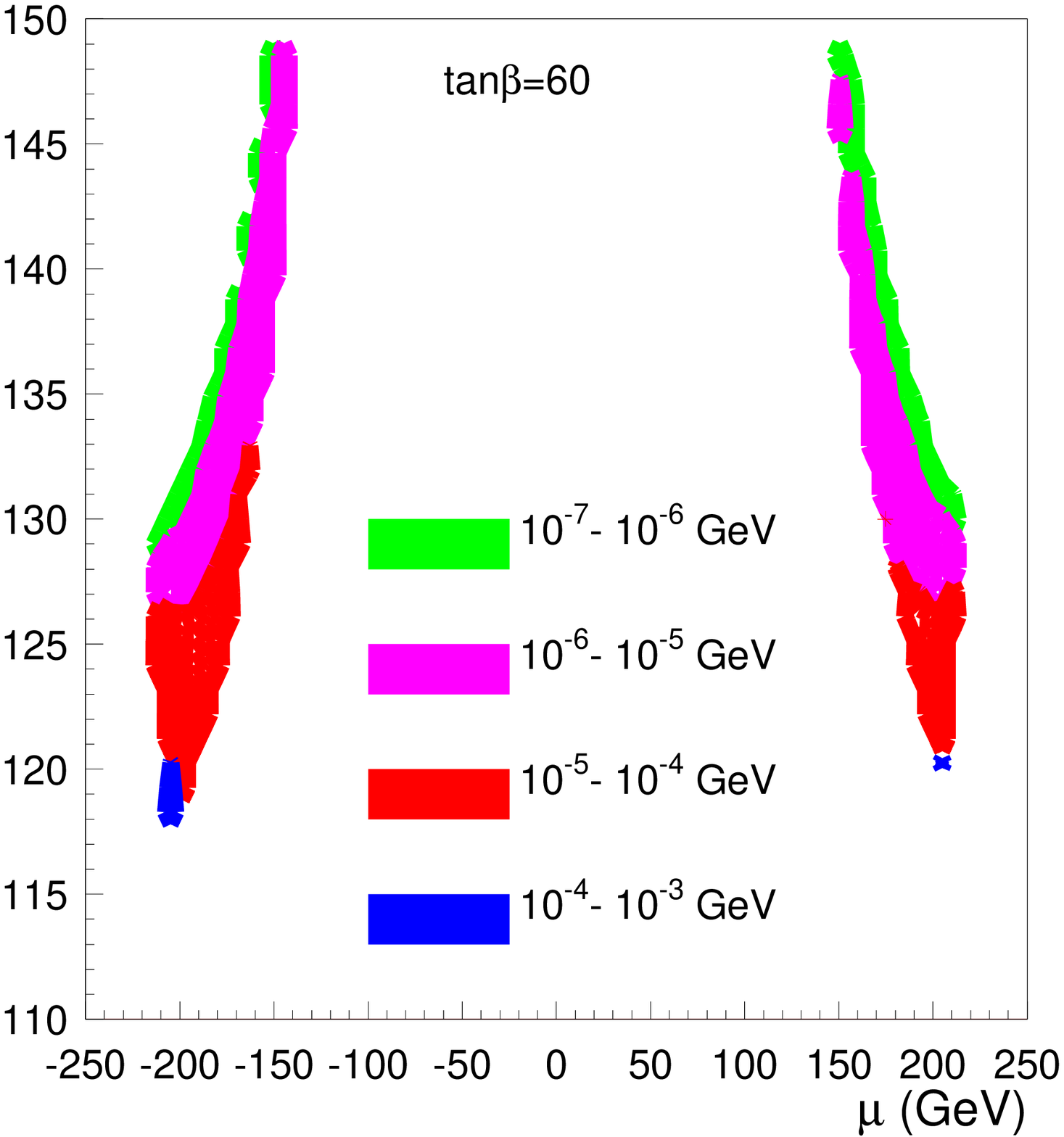,width=0.5\textwidth,height=0.5\textwidth}
\vskip -0.5cm
\caption{\label{fig2}:
{\em Three-body decay rates for $t \rightarrow \chi^+ \chi b$,
for $\tan\beta=$ 5, 60, subject to the LEP constraints. The
results are plotted in the $\mu, M_2$
plane, assuming gaugino-mass universality and $m_{\tilde{q}_{L,R}} =
250$~GeV.}}
\end{figure}

Bearing these points in mind, we present in Fig.~\ref{fig2}
contour plots for the decay width
for $t \rightarrow \chi^+ \chi b$, for $\tan \beta = 5, 60$.
We require $m_{\chi}+m_{\tilde{t}}>185$ GeV,
           $m_{\chi^+}+m_{\tilde{b}}>185$ GeV,
           $m_{\chi^+}>100$ GeV  and  $m_{\chi}>45$ GeV,
in order to keep 
the intermediate squarks at least 10~GeV off-shell, and
to be consistent with the present LEP~2 limits on chargino and neutralino
masses. For illustration,
we fix $m_{\tilde{q}_{L,R}} = 250$~GeV, noting, however, that mixing
effects can lead to smaller physical sfermion masses, especially
for large $\tan\beta$. 
Since we exclude the region of parameter space where
the physical squark masses are small enough to allow two-body
top decays, there is a sharp cut-off in the widths 
for large $\tan\beta$ and relatively light $m_{\tilde{q}_{L,R}}$,
visible in the second panel of Fig.~\ref{fig2}, where we
show our results for $\tan\beta = 60$.

\vspace*{0.2 cm}

\vspace*{0.2 cm}

Even for large $\tan\beta$, the contribution from the
Higgs-exchange diagram is at most 40-50$\%$ of the
total rate. As seen in Fig.~\ref{fig2}, generically we predict small
values of $\Gamma(t \rightarrow \chi^+ \chi b)$,
except for a part of the MSSM parameter
parameter space with relatively small $M_2$ and $\mu$,
close to the LEP limits, where
$10^{-6} \leq \Gamma(t \rightarrow \chi^+ \chi b) \leq 10^{-5}$ GeV.
This interesting region is larger for large $\tan\beta$ because,
for universal $m_{{\tilde q}_{L,R}}$, the lightest 
physical sbottom mass $m_{{\tilde b}_1}$ can be significantly
smaller than for small $\tan\beta$.
For larger values of $M_2$ and $\mu$, there is a kinematical cut-off,
since $m_{\chi^+} + m_\chi > m_{t}$, especially
for larger $\tan\beta$.

\begin{figure}[h]
\vspace*{-0.5cm}
{
\unitlength=1.5 pt
\SetScale{1.5}
\SetWidth{0.7}      
\begin{picture}(95,79)(0,0)
\Text(15.0,60.0)[r]{$t$}
\ArrowLine(16.0,60.0)(58.0,60.0) 
\Text(80.0,70.0)[l]{$\chi$}
\Line(58.0,60.0)(79.0,70.0) 
\Text(54.0,50.0)[r]{$\tilde{c}_1$}
\DashArrowLine(58.0,60.0)(58.0,40.0){1.0} 
\Text(80.0,50.0)[l]{$\chi^0_1$}
\Line(58.0,40.0)(79.0,50.0) 
\Text(80.0,30.0)[l]{$c$}
\ArrowLine(58.0,40.0)(79.0,30.0) 
\end{picture} \hspace*{-1cm} 
{} \qquad\allowbreak
\begin{picture}(95,79)(0,0)
\Text(15.0,60.0)[r]{$t$}
\ArrowLine(16.0,60.0)(58.0,60.0) 
\Text(80.0,70.0)[l]{$\chi$}
\Line(58.0,60.0)(79.0,70.0) 
\Text(54.0,50.0)[r]{$\tilde{c}_2$}
\DashArrowLine(58.0,60.0)(58.0,40.0){1.0} 
\Text(80.0,50.0)[l]{$\chi$}
\Line(58.0,40.0)(79.0,50.0) 
\Text(80.0,30.0)[l]{$c$}
\ArrowLine(58.0,40.0)(79.0,30.0) 
\end{picture} 
\vskip -1.5cm 
\begin{picture}(95,79)(0,0)
\Text(15.0,60.0)[r]{$t$}
\ArrowLine(16.0,60.0)(58.0,60.0) 
\Text(80.0,70.0)[l]{$\chi$}
\Line(58.0,60.0)(79.0,70.0) 
\Text(54.0,50.0)[r]{$\tilde{t}_1$}
\DashArrowLine(58.0,60.0)(58.0,40.0){1.0} 
\Text(80.0,50.0)[l]{$\chi$}
\Line(58.0,40.0)(79.0,50.0) 
\Text(80.0,30.0)[l]{$c$}
\ArrowLine(58.0,40.0)(79.0,30.0) 
\end{picture}   \hspace*{-1cm}
{} \qquad\allowbreak
\begin{picture}(95,79)(0,0)
\Text(15.0,60.0)[r]{$t$}
\ArrowLine(16.0,60.0)(58.0,60.0) 
\Text(80.0,70.0)[l]{$\chi$}
\Line(58.0,60.0)(79.0,70.0) 
\Text(54.0,50.0)[r]{$\tilde{t}_2$}
\DashArrowLine(58.0,60.0)(58.0,40.0){1.0} 
\Text(80.0,50.0)[l]{$\chi$}
\Line(58.0,40.0)(79.0,50.0) 
\Text(80.0,30.0)[l]{$c$}
\ArrowLine(58.0,40.0)(79.0,30.0) 
\end{picture} \ 
}
\vspace*{-1.5cm}
\caption{\label{fig3}: {\em Three-body top decays pairs of neutralinos
in the MSSM: $t \rightarrow \chi \chi c$, in the presence of
$\tilde{c}-\tilde{t}$ mixing.}}
\end{figure}
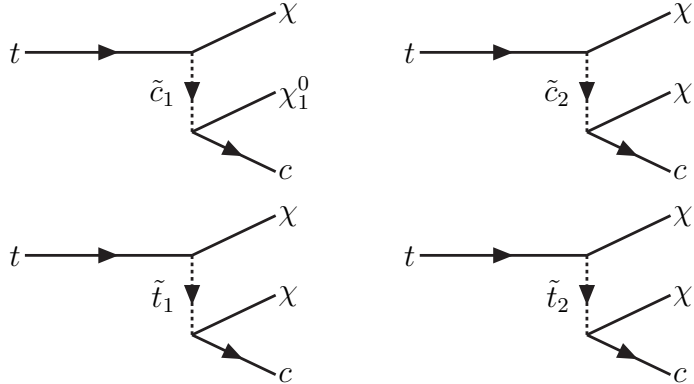

In the case of non-negligible $\tilde{c}-\tilde{t}$ mixing,
one can in principle produce via the diagrams shown in Fig.~\ref{fig3} a
pair of neutralinos: $t \rightarrow \chi \chi c$,
instead of a neutralino and a chargino: $t \rightarrow \chi^+ \chi b$.
However, although the phase-space suppression of the process is smaller, 
(i) squark-flavour mixing, and  (ii) the rather small couplings
of neutralinos to squarks (even for maximal mixing)
in the part of the MSSM parameter space
that is relevant for our calculation, both drive the
expected rates to very small values. Even for large
$\tilde{c}-\tilde{t}$ mixing, the expected partial decay width
is typically below $10^{-7}-10^{-8}$~GeV, and hence unobservable
at either the LHC or TESLA. We therefore do not present detailed
results for this decay mode.

\section{$R$-violating Top Decays}

The phenomenology of supersymmetry
may not be restricted to the MSSM: the most general
$SU(3)_c\times SU(2)_L\times U(1)_Y$-invariant superpotential with the
field content of the minimal supersymmetric extension of the Standard
Model also contains the terms 
\beq
W=\lam_{ijk} L_iL_j{\bar E}_k+\lam'_{ijk}L_iQ_j{\bar D_k}+
\lam''_{ijk}{\bar U_i}{\bar D_j}{\bar D_k}
\label{eq:superpot}
\eeq
where $L$ $(Q)$ are the left-handed lepton (quark) superfields, and 
${\bar
E},{\bar D}$ $({\bar U})$ are the corresponding left-handed
antilepton (antiquark) superfields. The
first two operators violate lepton number, whilst the third violates
baryon
number. If both lepton- and baryon-number-violating operators  were
present at
the same time in the low-energy Lagrangian, they would lead to
unacceptably
fast proton decay. However, it has been shown~\cite{sym} that there exist
symmetries which allow the violation of only a subset of these operators,
leading to very rich phenomenology~\cite{phen}, while being
consistent with the limits on proton decay. 

There are $45$ $R$-violating operators consistent with
$SU(2)$ and $SU(3)$ invariance,
$36$ associated with lepton-number 
violation and $9$ with baryon-number violation. 
Amongst all these, those operators involving a top quark
can be studied directly in top decays, whereas they are
currently only constrained weakly by indirect arguments.
Since the $\bar{U}\bar{D}\bar{D}$ operators are likely to be
`drowned' by the QCD background, we restrict 
our attention to $L_iQ_3\bar{D}_k$ and
$L_iQ_j\bar{D}_3$ operators, which are
bounded by several processes,
as summarized in the Tables~\cite{constraints}.
In most cases, these bounds cases simply 
scale with the squark masses, so
we may allow for significantly larger couplings if
larger sfermion masses are involved in these processes.
In view of
the very weak bounds on the relevant $R$-violating
operators, either rare top decays will
be observed, or certain
bounds can be tightened. 

\begin{table}[h]
\begin{center}
\begin{tabular}{|lll||lll||lll|}
\hline
$ijk$ & $\lambda'_{ijk}$ & Sources & $ijk$ & $\lambda'_{ijk}$ &
Sources & $ijk$ & $\lambda'_{ijk}$ & Sources 
 \\ \hline
131 & 0.035 & A.P.P.V. & 231 & 0.22 & $\nu_\mu$ D.I.S. & 
331 & 0.48 & $R_\tau$ \\
132 & 0.34 & $R_e$ & 232 & 0.36 & $R_\mu$ & 332 & 
0.48 & $R_\tau$ \\
133 & 0.0007 & $\nu_e$ mass & 233 & 0.36 & $R_\mu$ & 333 & 
0.48 & $R_\tau$ \\
\hline
\end{tabular}
\caption{\it 
Upper limits on individual $LQ\bar{D}$ operators involving
the top quark~\cite{constraints}, 
assuming $m_{\tilde{f}} = 100$ GeV. The allowed couplings
increase as the sfermion masses become higher. We denote 
atomic physics parity violation by A.P.P.V., deep-inelastic
scattering by D.I.S., and $Z^0$ decay branching ratios into
$\tilde \ell \ell$ measured at LEP by $R_{\ell}$.}

\vspace*{0.9 cm}

\begin{tabular}{|lll||lll|}
\hline
Combinations & Limits & Sources & Combinations & Limits & Sources \\
\hline
$\lambda'_{i13} \lambda'_{i31}$ & $8. 10^{-8}$ & $\Delta m_B$ &
$\lambda'_{1k1} \lambda'_{2k2}$ & $8. 10^{-7}$ & $K_L \rightarrow \mu
e$ \\
$\lambda'_{1k1} \lambda'_{2k1}$ & $5. 10^{-8}$ & $\mu {\rm Ti} 
\rightarrow e {\rm Ti}$ & 
$\lambda_{231} \lambda_{131}$ & $7. 10^{-7}$ & $\mu \rightarrow 3e$ \\
\hline
\end{tabular}
\caption{\it 
Upper limits on 
some important products of $R$-violating couplings involving
the top quark~\cite{constraints}, assuming $m_{\tilde{f}} = 100$ GeV.
These  limits are shown for the sake of completeness: the $t$-decay
processes we consider later are sensitive to different products of
couplings.}
\end{center}
\end{table}

In the presence of $R$-violating interactions,
decays of the
top quark into a {\em single sparticle} become possible, as do
other potential signatures.

$\bullet$ If there is at least one light sfermion with
$m_{\tilde{f}} < m_{t}$, we can expect two-body top decays
$t \rightarrow \tilde{f} f'$. These have already been studied in the
literature~\cite{2body1}. However, such a
possibility is rather disfavoured, in view of the current experimental
bounds and theoretical expectations, and we do not discuss this
possibility further here.

$\bullet$ If the sfermion is a squark, one
may also produce single charginos and neutralinos
in three-body final states. Since there is only one massive
particle in the final state,
the phase space suppression is potentially less
severe than in the three-body MSSM decays to chargino/neutralino
pairs discussed earlier.

$\bullet$ There may also be decays of the top quark to three 
conventional fermions, via two insertions of $R$-violating operators.

In the remainder of this paper, we discuss the two latter options above.

$\bullet$ Finally, in the case of small $R$-violating couplings,
the dominant supersymmetric decay mode of the $t$ quark would tend to be
$\chi \chi^+b$, whilst the chargino
and neutralino woul subsequently decay to fermions via
the $R$-violating interactions.
Since we predict relatively small widths
for such MSSM top decays, we do not discuss this issue
in detail.

\subsection{Top Decay into a Single Neutralino}

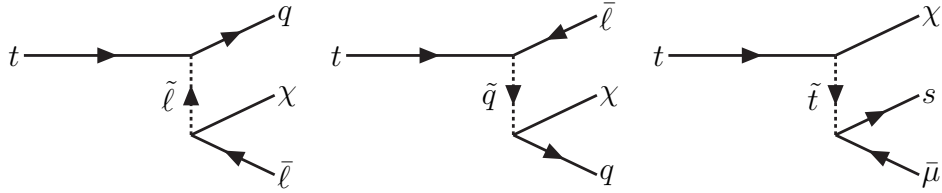
\begin{figure}[h]
{
\unitlength=1.5 pt
\SetScale{1.5}
\SetWidth{0.7}      
\begin{picture}(95,79)(0,0)
\Text(15.0,60.0)[r]{$t$}
\ArrowLine(16.0,60.0)(58.0,60.0) 
\Text(80.0,70.0)[l]{$q$}
\ArrowLine(58.0,60.0)(79.0,70.0) 
\Text(54.0,50.0)[r]{$\tilde{\ell}$}
\DashArrowLine(58.0,40.0)(58.0,60.0){1.0} 
\Text(80.0,50.0)[l]{$\chi$}
\Line(58.0,40.0)(79.0,50.0) 
\Text(80.0,30.0)[l]{$\bar{\ell}$}
\ArrowLine(79.0,30.0)(58.0,40.0) 
\end{picture} \hspace*{-1cm}
\begin{picture}(95,79)(0,0)
\Text(15.0,60.0)[r]{$t$}
\ArrowLine(16.0,60.0)(58.0,60.0) 
\Text(80.0,70.0)[l]{$\bar{\ell}$}
\ArrowLine(79.0,70.0)(58.0,60.0) 
\Text(54.0,50.0)[r]{$\tilde{q}$}
\DashArrowLine(58.0,60.0)(58.0,40.0){1.0} 
\Text(80.0,50.0)[l]{$\chi$}
\Line(58.0,40.0)(79.0,50.0) 
\Text(80.0,30.0)[l]{$q$}
\ArrowLine(58.0,40.0)(79.0,30.0) 
\end{picture}  \hspace*{-1cm}
\begin{picture}(95,79)(0,0)
\Text(15.0,60.0)[r]{$t$}
\ArrowLine(16.0,60.0)(58.0,60.0) 
\Text(80.0,70.0)[l]{$\chi$}
\Line(58.0,60.0)(79.0,70.0) 
\Text(54.0,50.0)[r]{$\tilde{t}$}
\DashArrowLine(58.0,60.0)(58.0,40.0){1.0} 
\Text(80.0,50.0)[l]{$s$}
\ArrowLine(58.0,40.0)(79.0,50.0) 
\Text(80.0,30.0)[l]{$\bar{\mu}$}
\ArrowLine(79.0,30.0)(58.0,40.0) 
\end{picture} \ 
}
\caption{\label{fig4}{\em Three-body $t$-quark decays to a single
neutralino, via $R$-violating operators $L_i Q_3 \bar{D}_k$.}}
\end{figure}

Top decays into a single neutralino and two conventional fermions has been
discussed previously in~\cite{3bodyneu},
and may proceed via the diagrams shown in Fig.~\ref{fig4}.

\begin{figure}[h]
\vskip -0.5cm
\hspace*{-0.5cm}\epsfig{file=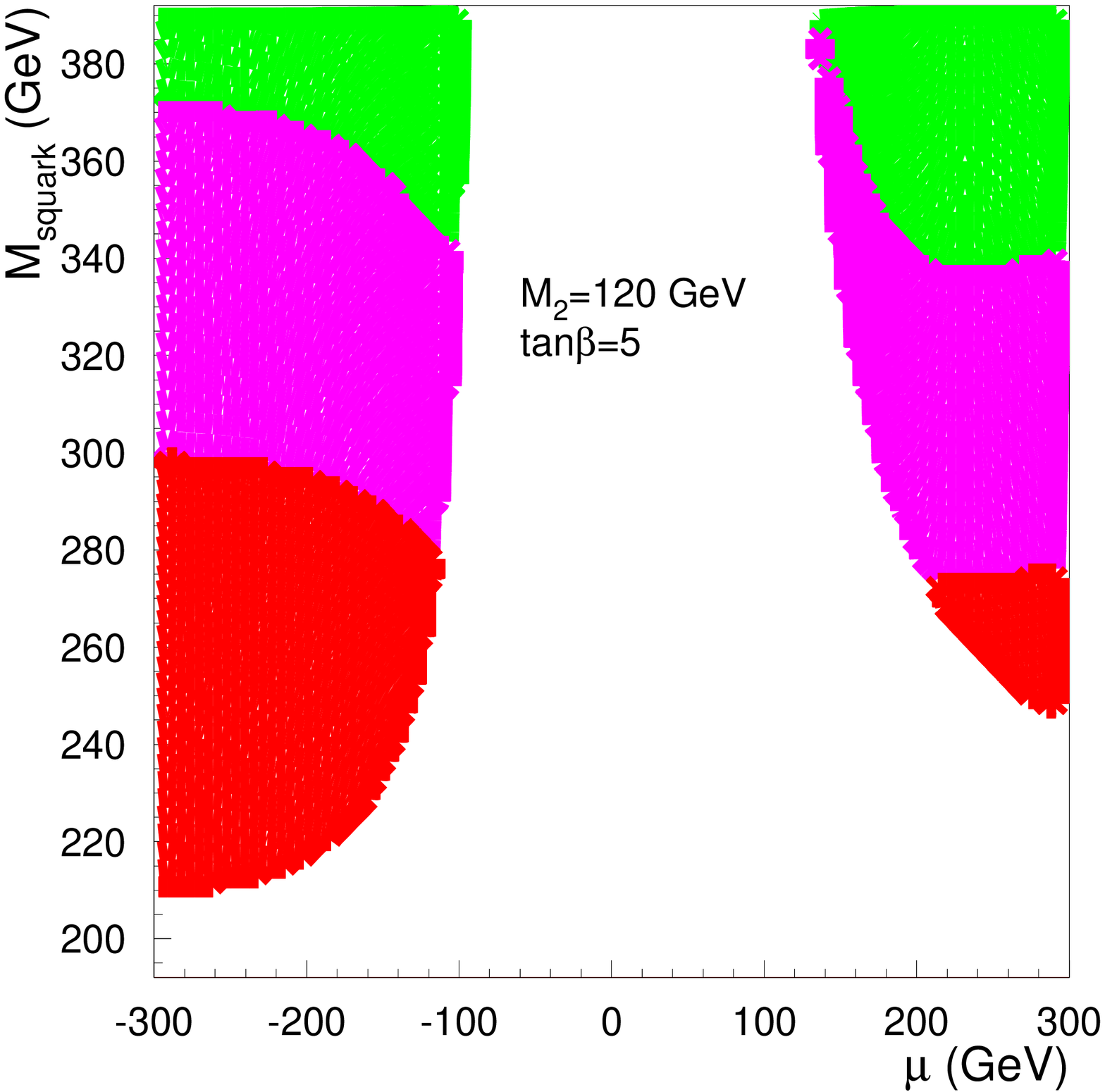,width=0.5\textwidth,height=0.5\textwidth}
\hspace*{-0.5cm}\epsfig{file=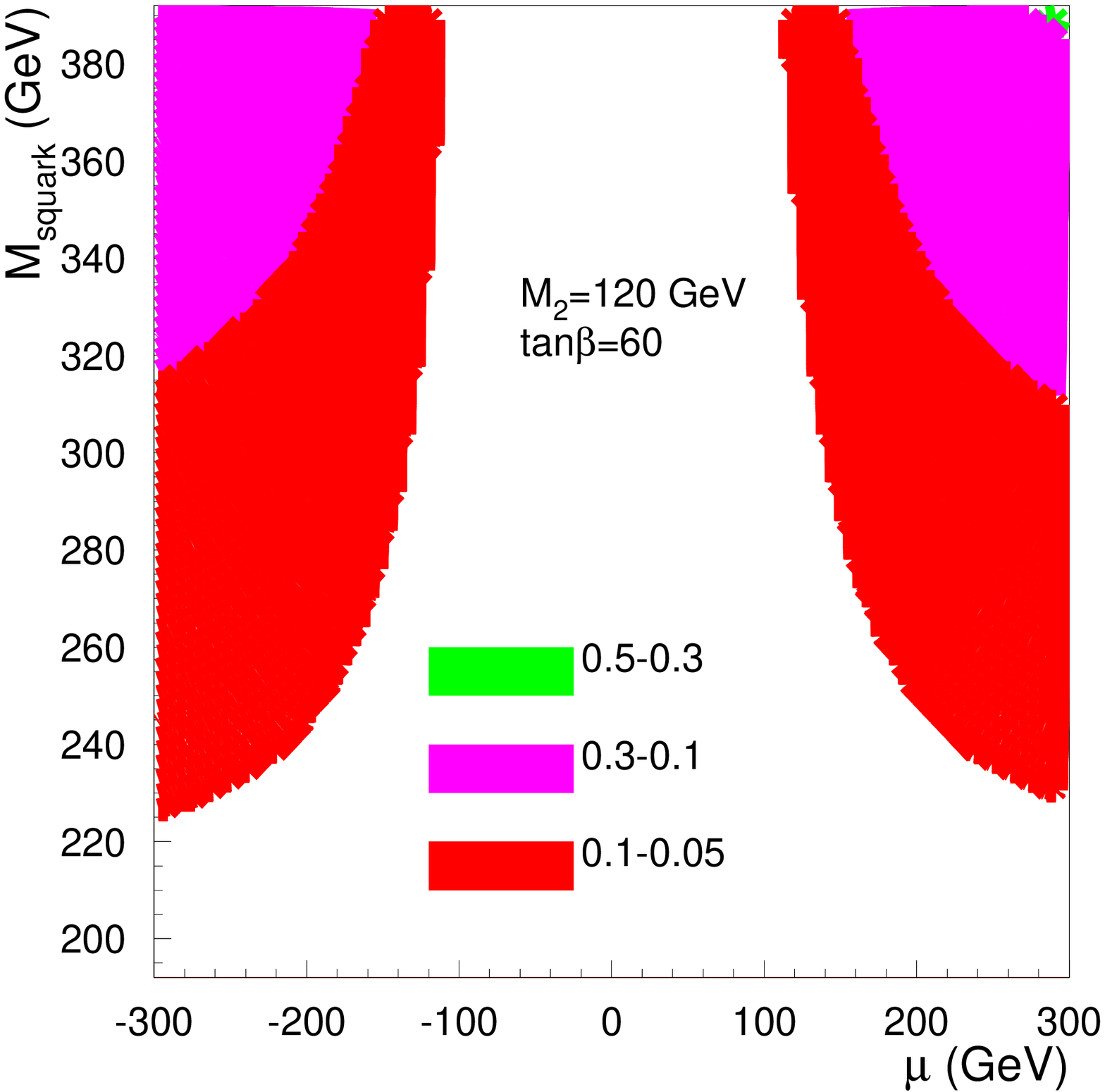,width=0.5\textwidth,height=0.5\textwidth}
\vskip -0.5cm
{\bf \caption{\label{fig5}
{\em Bounds on $R$-violating couplings available from
$t$-quark decays to single neutralinos,
for $\tan\beta=$ 5, 60. The different contours
denote $\lambda'$ in the ranges
0.1-0.3,0.3-0.5 and 0.5-1, for a top
width of $10^{-6}$ GeV.}}}
\end{figure}

\noindent
The possible signal from
this decay mode of the $t$ quark is striking, due to the
likely subsequent decay of the neutralino
to an $R$-even final state such as $ {\tilde\chi} \ra q \bar{q}' \ell$
or ${\tilde\chi} \ra q \bar{q}' \nu$.
In particular, the subsequent decays of neutralinos can
give rise to like-sign lepton signals, even if there is
only one dominant R-violating
operator \cite{LMc}. This is because the neutralino
is a Majorana spinor, and therefore can decay equally
into leptons and antileptons.
Any like-sign dilepton final state would be
exotic, with no Standard Model physics background.
Non-observation of single neutralino production in $t$ decay at the
LHC would strengthen the above bounds on $R$-violating couplings
as a function of the other supersymmetric model parameters,
as illustrated in Fig.~\ref{fig5}, where we assume universal
sfermion masses, shown as $m_{squark}$ on the vertical axis.

\subsection{Top Decay into a Single Chargino}

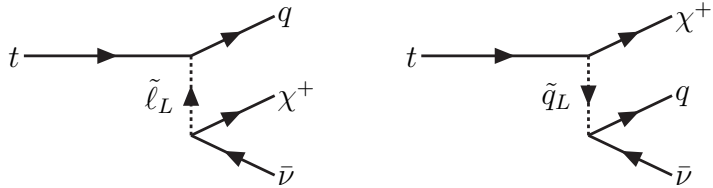
\begin{figure}[h]
{
\unitlength=1.5 pt
\SetScale{1.5}
\SetWidth{0.7}      
\begin{picture}(95,79)(0,0)
\Text(15.0,60.0)[r]{$t$}
\ArrowLine(16.0,60.0)(58.0,60.0) 
\Text(80.0,70.0)[l]{$q$}
\ArrowLine(58.0,60.0)(79.0,70.0) 
\Text(54.0,50.0)[r]{$\tilde{\ell}_L$}
\DashArrowLine(58.0,40.0)(58.0,60.0){1.0} 
\Text(80.0,50.0)[l]{$\chi^+$}
\ArrowLine(58.0,40.0)(79.0,50.0) 
\Text(80.0,30.0)[l]{$\bar{\nu}$}
\ArrowLine(79.0,30.0)(58.0,40.0) 
\end{picture} \ 
\begin{picture}(95,79)(0,0)
\Text(15.0,60.0)[r]{$t$}
\ArrowLine(16.0,60.0)(58.0,60.0) 
\Text(80.0,70.0)[l]{$\chi^+$}
\ArrowLine(58.0,60.0)(79.0,70.0) 
\Text(54.0,50.0)[r]{$\tilde{q}_L$}
\DashArrowLine(58.0,60.0)(58.0,40.0){1.0} 
\Text(80.0,50.0)[l]{$q$}
\ArrowLine(58.0,40.0)(79.0,50.0) 
\Text(80.0,30.0)[l]{$\bar{\nu}$}
\ArrowLine(79.0,30.0)(58.0,40.0) 
\end{picture}\
 }
\caption{\label{fig6} {\em Three-body $t$-quark decays to a single chargino,
via $R$-violating operators $L_i Q_3 \bar{D}_k$.}}
\end{figure}

Top-quark decay into a single chargino and two conventional fermions
may proceed via the diagrams shown in Fig.~\ref{fig6}.

\noindent
Note that we show in Fig.~\ref{fig6} diagrams with an
electroweak-doublet field in the propagator, since if this were not the
case the wino component of the chargino
would be decoupled, and the chargino vertex would involve
only the relatively suppressed higgsino coupling. However, even in this
relatively favourable case, the decay width is suppressed by a 
significant factor compared to the neutralino case. This is because
the chargino vertex involves a $\tilde{d}_L$ or a 
$\tilde{\ell}_L$ state, instead of a $\tilde{u}_L$, and thus
is proportional to the $U_{ij}$ mixing matrix element, 
in the notation of Gunion and Haber, rather than the
$V_{ij}$ mixing element.

\begin{figure}[h]
\vskip -0.5cm
\hspace*{-0.5cm}\epsfig{file=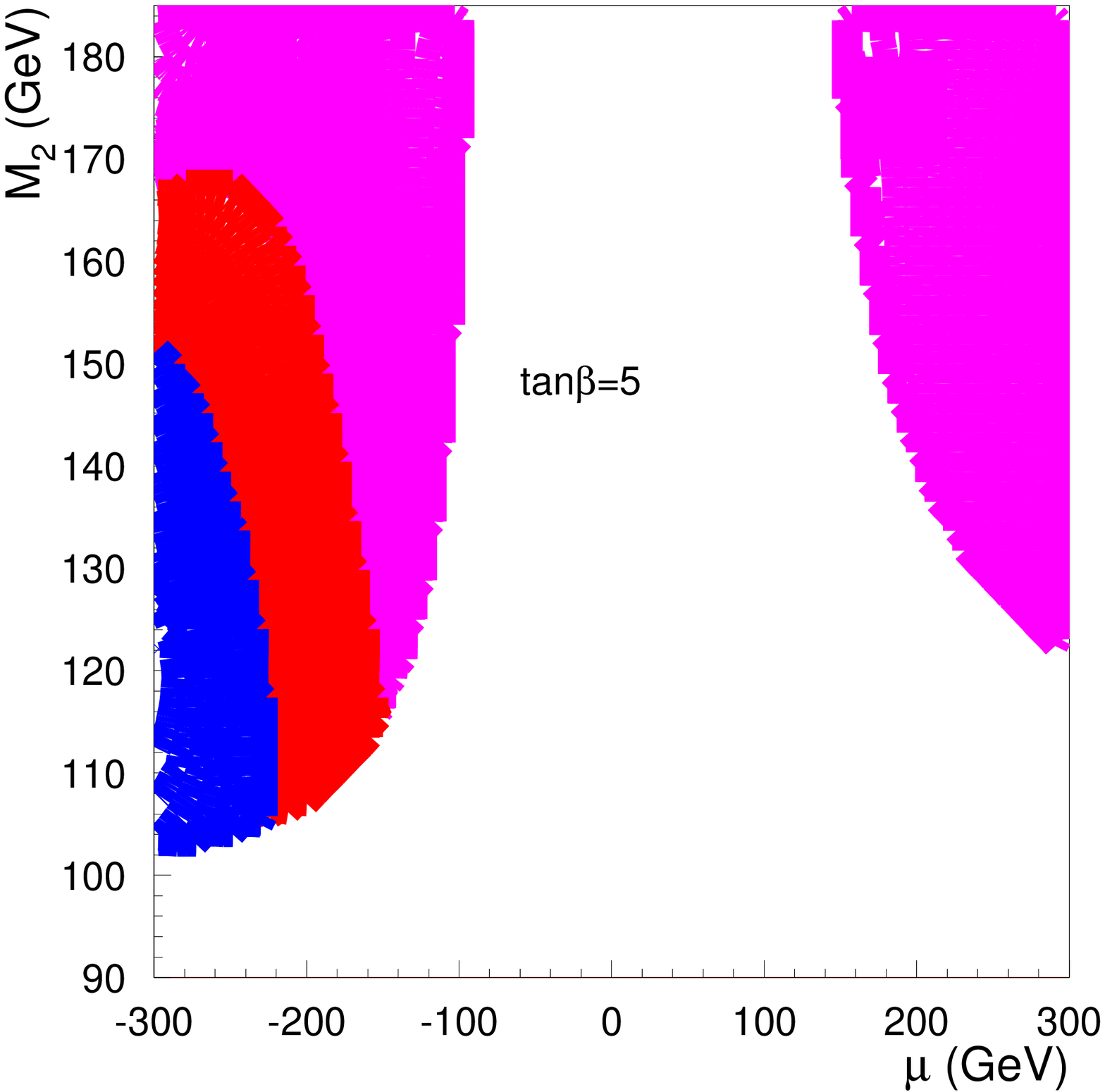,width=0.5\textwidth,height=0.5\textwidth}
\hspace*{-0.5cm}\epsfig{file=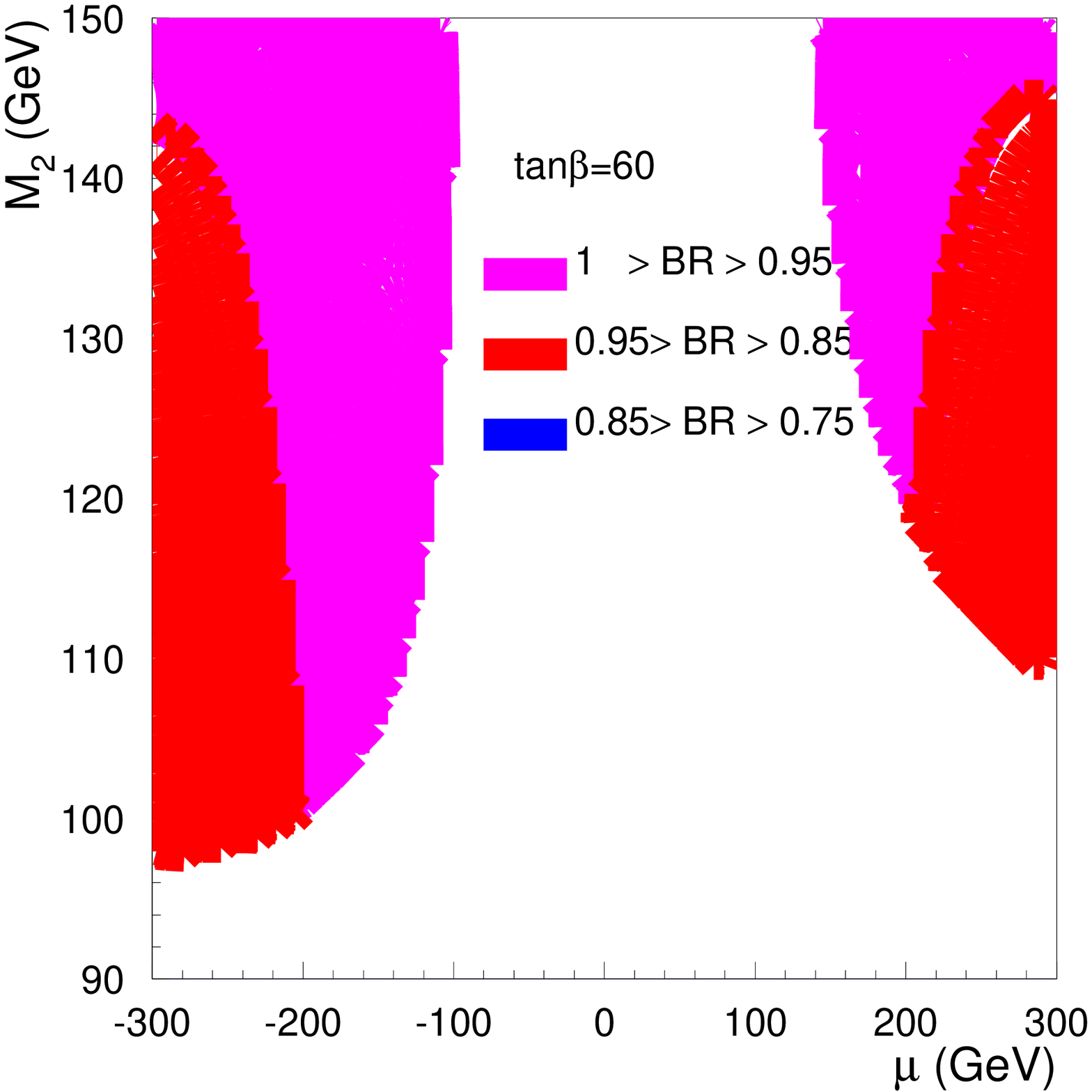,width=0.5\textwidth,height=0.5\textwidth}
\vskip -0.5cm
\caption{\label{fig7}
{\em Contours of the branching ratio for
MSSM chargino decays, for $\tan\beta=$ 5, 60,
$\lambda' = 0.2$ 
and $m_{\tilde{f}} = 350$ GeV.
}}
\end{figure}

We should also stress that charginos have two possible important
decay modes: cascade decay via the lightest neutralino,
and direct decay via $R$-violating coupling(s)~\cite{hmp}.
For instance, for the lightest chargino we have
the $R$-conserving channel ${\chi}^-\ra {\chi} + (W^-)^* \ra
{\chi} + f{\bar f}'$,
where $f{\bar f}'$ are the decay products of the $W$ boson,
which may be real or virtual, depending on the mass
gap between the lightest chargino and neutralino,
or the $R$-violating channels ${\chi}^-_1\ra q \bar{q}' \ell$,
${\chi}^-\ra q \bar{q}' \nu$,
where the flavours of the quarks and the leptons
depend on the flavour structure of the $R$-violating coupling.
Which of the two processes will dominate clearly depends on
(i) the strength of the $R$-violating operator:
the stronger the operator, the larger the decay rate for 
direct decay of the chargino, and
(ii) the difference in mass between the chargino and neutralino:
if the mass gap between the two states is very small,
the cascade decay is suppressed by phase space. 
We investigate these issues in the contour plots that appear
in Fig.~\ref{fig7}.

\begin{figure}[h]
\vskip -0.5cm
\hspace*{-0.5cm}\epsfig{file=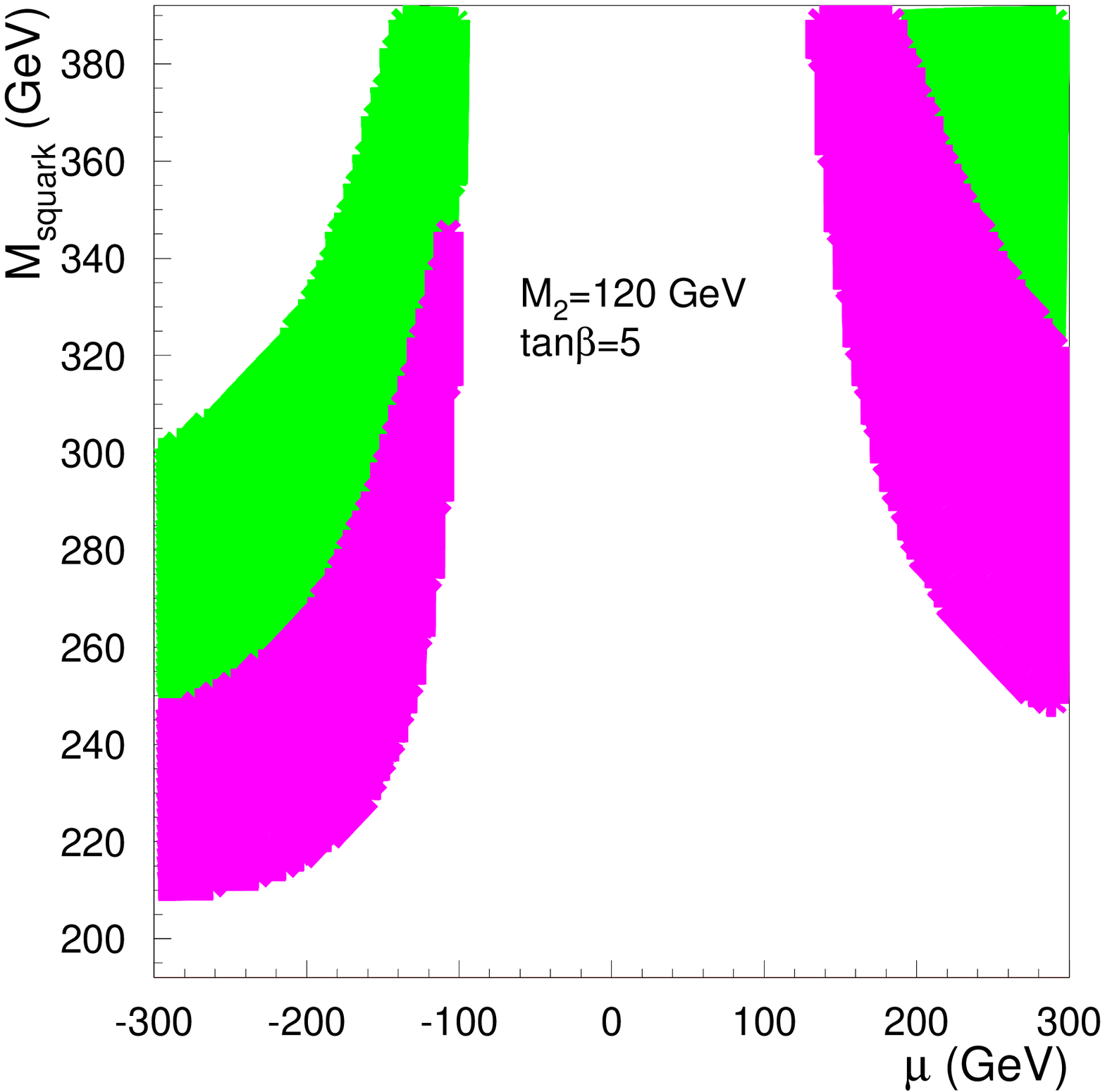,width=0.5\textwidth,height=0.5\textwidth}
\hspace*{-0.5cm}\epsfig{file=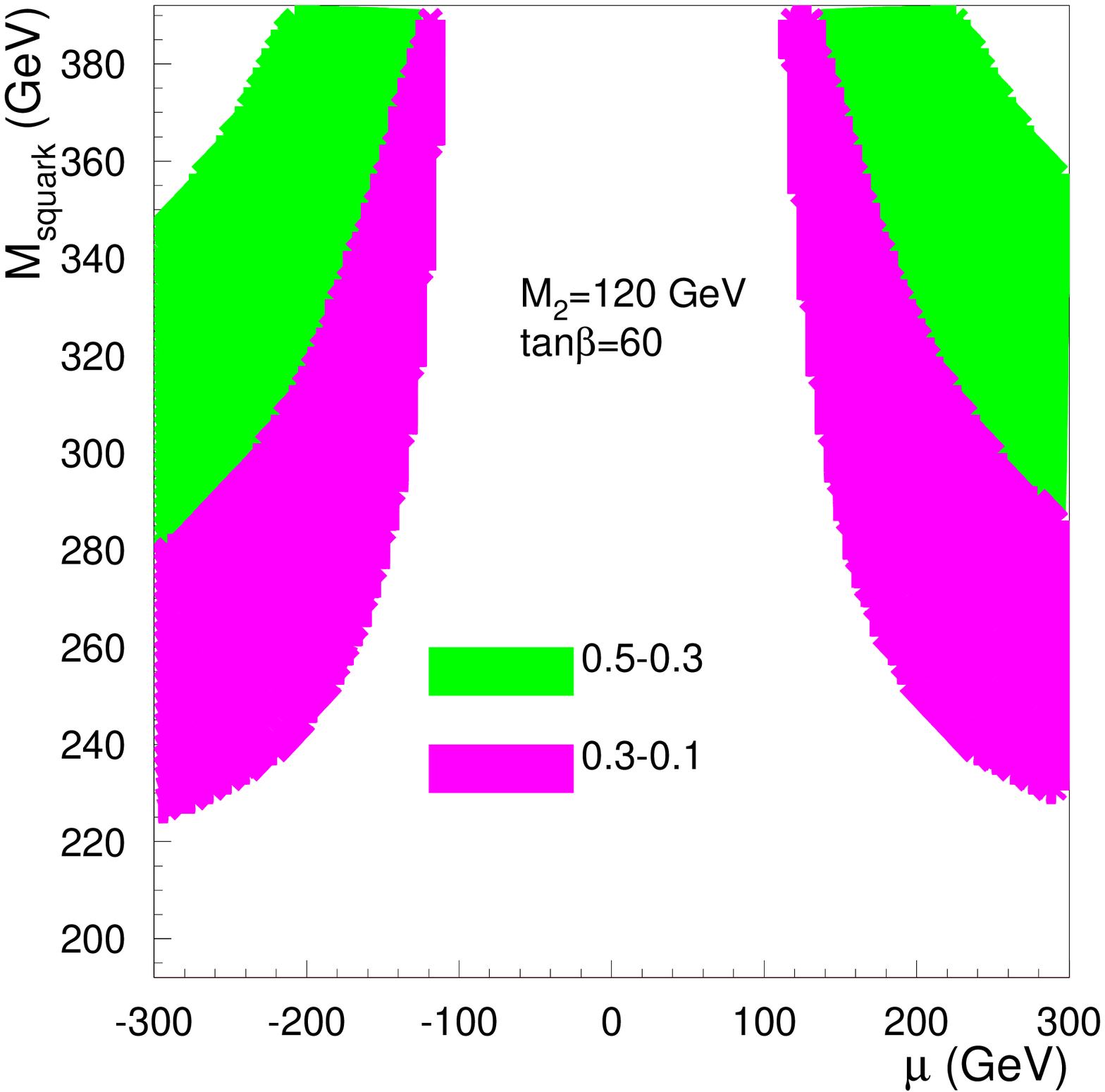,width=0.5\textwidth,height=0.5\textwidth}
\vskip -0.5cm
\caption{\label{fig8}
{\em Bounds on $R$-violating couplings available from
$t$-quark decays to single charginos,
for $\tan\beta=$ 5, 60. The different contours
denote $\lambda'$ in the ranges
0.1-0.3,0.3-0.5 and 0.5-1, for a partial top decay
width of $10^{-6}$ GeV.}}
\end{figure}

As seen in Fig.~\ref{fig7}, cascade chargino decays 
through neutralinos, which
give rise to like-sign dilepton signals, dominate
over the $R$-violating direct chargino decay.
This situation is slightly altered for
smaller sfermion masses, but the $R$-conserving MSSM
chargino to neutralino decay still dominates.
The contour plots in Fig.~\ref{fig8} indicate that
the coupling bounds from $R$-violating top decays to charginos
are likely to be weaker than those from neutralinos.
Nevertheless, the chargino decay width is 
significant, and may provide a complementary channel
for testing models of any $R$-violating top decays seen.

\subsection{Top Decays into Three Conventional Fermions}

We assume that $t$ decay occurs initially via
a single dominant $R$-violating operator.
This has to involve the top quark, and must
therefore be of the type $L_iQ_3 \bar{D}_k$.
Then, since the $SU(2)$-singlet quark
field is necessarily used in the
propagator, the final states are specified to be
$t \rightarrow \bar{\ell}_i \nu_i b$.
This process is likely to be drowned in the Standard Model
background due to the $W$-exchange diagram, and therefore is
lihely to be invisible, although there might be some hope in
a detailed comparison of the different 
$t \rightarrow \bar{\ell}_i \nu_i b$ branching modes.

\begin{figure}[h]
{
\unitlength=1.5 pt
\SetScale{1.5}
\SetWidth{0.7}      
\begin{picture}(95,79)(0,0)
\Text(15.0,60.0)[r]{$t$}
\ArrowLine(16.0,60.0)(58.0,60.0) 
\Text(80.0,70.0)[l]{$e^+,\bar{\mu},\bar{\tau}$}
\ArrowLine(79.0,70.0)(58.0,60.0) 
\Text(54.0,50.0)[r]{$\tilde{d}_R$}
\DashArrowLine(58.0,60.0)(58.0,40.0){1.0} 
\Text(80.0,50.0)[l]{$d,s,b $}
\ArrowLine(58.0,40.0)(79.0,50.0) 
\Text(80.0,30.0)[l]{$\nu$}
\ArrowLine(58.0,40.0)(79.0,30.0) 
\end{picture} \hspace*{0.3cm}
\begin{picture}(95,79)(0,0)
\Text(15.0,60.0)[r]{$t$}
\ArrowLine(16.0,60.0)(58.0,60.0) 
\Text(80.0,70.0)[l]{$e^+,\bar{\mu},\bar{\tau}$}
\ArrowLine(79.0,70.0)(58.0,60.0) 
\Text(54.0,50.0)[r]{$\tilde{d}_R$}
\DashArrowLine(58.0,60.0)(58.0,40.0){1.0} 
\Text(80.0,50.0)[l]{$u,c$}
\ArrowLine(58.0,40.0)(79.0,50.0) 
\Text(80.0,30.0)[l]{$e^-,\mu,\tau $}
\ArrowLine(58.0,40.0)(79.0,30.0) 
\end{picture} \ 
}
\caption{\label{fig9}{\em Three-body $t$-quark decays to 
conventional fermions, via the $R$-violating operators
$L_i Q_3 \bar{D}_1$  and $L_j Q_2 \bar{D}_1$.}}
\end{figure}
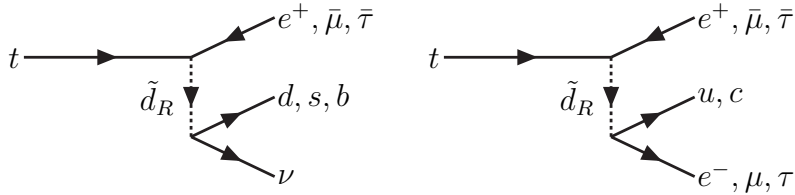

The situation changes when two $R$-violating operators
are simultaneously large. The first should be of the
type $L_iQ_3 \bar{D}_k$, whilst the second should be of
the type $L_sQ_q\bar{D}_r$, where $r = k$~\footnote{There
might also be
combinations of $L_iQ_3 \bar{D}_k$ and $LL\bar{E}$ operators.
However, in view of the very strong bounds on
$LL\bar{E}$ couplings~\cite{constraints}, we do not consider
this possibility further.}. Since the $SU(2)$-singlet
squark in the propagator can lead to either an up-type 
quark and
a charged lepton or a down-type quark and a neutrino
(where $q \neq 3$ for the second operator),
in 50$\%$ of the cases we observe top decay to a
quark and two charged leptons, via the second diagram of
Fig.~\ref{fig9}. On the other hand, for  $q=3$ the final
state is necessarily
$t \rightarrow \bar{\ell}_i \nu_i b$, which we consider
to be indistinguishable from the Standard Model background.

\begin{figure}[h]
\vskip -0.5cm
\hspace*{0.5cm}
\epsfig{file=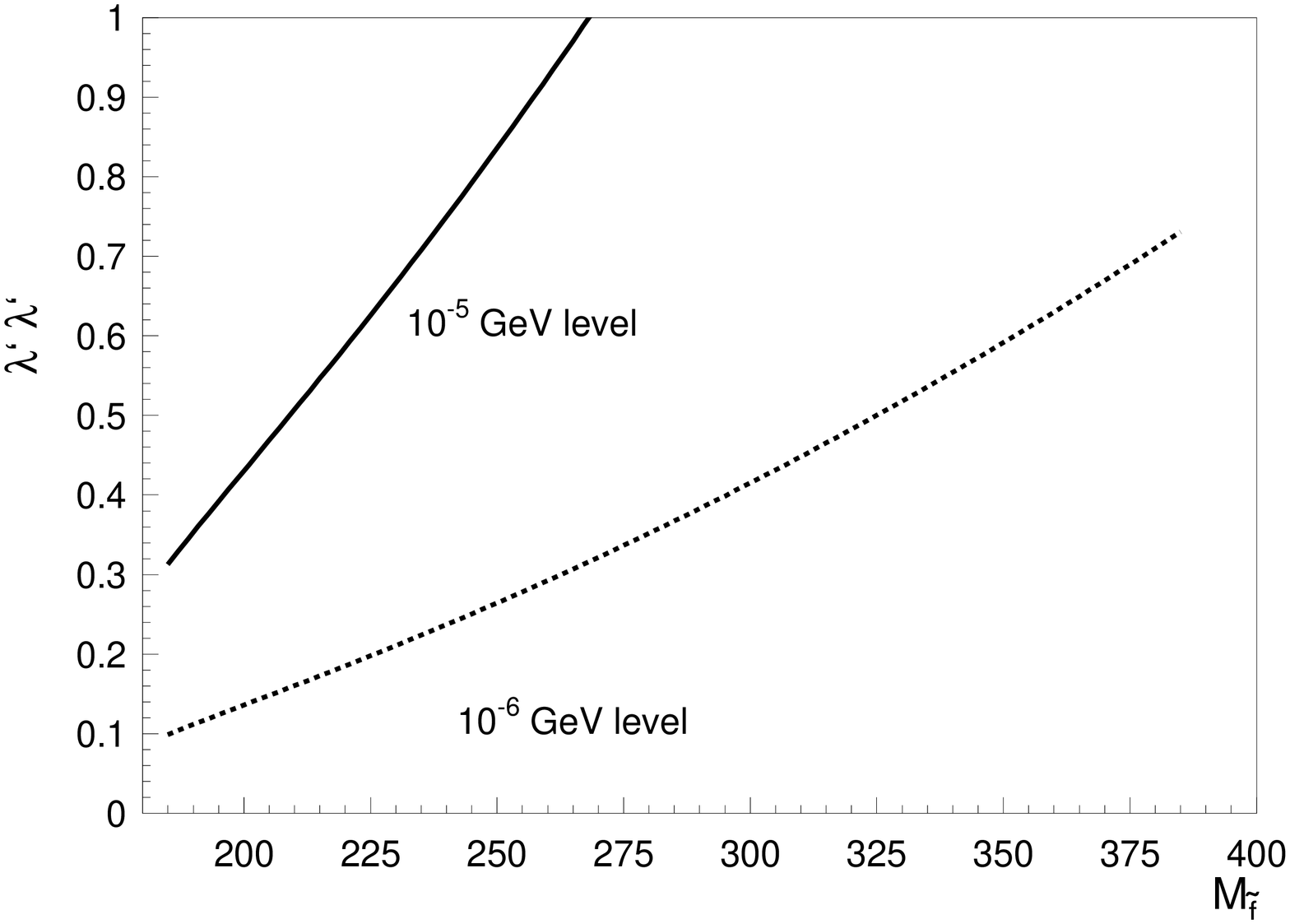,width=0.9\textwidth,height=0.7\textwidth}
\vskip -0.5cm
\caption{\label{fig10}
{\em Bounds on 
$\lambda'_{i3k} \lambda'_{spk}$
as a function of $m_{\tilde{f}}$ for
a partial decay width discovered at the level of
$10^{-5}$ or $10^{-6}$ GeV.}}
\end{figure}

\vspace*{0.2 cm}

\noindent
We find that, for $\lambda'_{ijk} \lambda'_{spk} \geq
0.09$ and squark masses of $200$ GeV, 
the width for this three-body decay is ${\cal O}(10^{-5})$ GeV.
Given that, for
certain flavours, the current bound
on $\lambda'_{ijk} \lambda'_{spk}$ for sfermions of mass 200~GeV
can be as large as unity, $R$-violating top decays to three
conventional fermions are promising.
The absence of a signal could again be translated to a bound
on the respective products of $R$-violating couplings, as seen
in Fig.~\ref{fig10}.


\section{Conclusions}

We have studied three-body supersymmetric top decays
to charginos, neutralinos and fermions,
both in the MSSM and in schemes with $R$-violation.
Whilst the MSSM top decay widths are typically below
$10^{-5}$ GeV, the $R$-violating decays
to single charginos, single neutralinos, and all-fermion
final states can be larger, for 
values of the coupling constants of $R$-violating interactions
consistent with the present experimental upper limits.

The cascade decays of the charginos and neutralinos would
lead to signals with like-sign lepton events, whilst the
top decays to three fermions including two charged leptons
are also signatures that may be of interest.
Moreover, the decay widths of the processes
studied are highly correlated and very
sensitive to the supersymmetric model parameters.
This means that, if such signals are observed, 
the allowed ranges of these parameters will be 
severely constrained.
In the absence of signals, new bounds on single or
products of $R$-violating couplings may be derived.

We conclude that three-body supersymmetric top decays
may provide a fertile testing-ground for probing
supersymmetric extensions of the Standard Model, particularly
extensions of the MSSM to include $R$-violating couplings.

\vspace*{0.2cm}
\begin{center}
{\bf Acknowledgements}
\end{center}
\vspace*{-0.3cm}
We are happy to acknowledge the use of the CompHEP package~\cite{
COMPHEP} to evaluate the
Feynman diagrams for the processes studied in this paper.

\pagebreak

\end{document}